\documentclass[a4paper,12pt]{article}  
\usepackage{latexsym}  
\usepackage{amsfonts} 
\usepackage{multirow} 
\usepackage{array} 
\usepackage{tabularx}  
\usepackage{amsmath}  
\usepackage[dvips]{graphicx} 
\usepackage{epsfig} 
\newcommand{\be}{\begin{equation}}
\newcommand{\ee}{\end{equation}}
\newcommand{\bea}{\begin{eqnarray}}
\newcommand{\eea}{\end{eqnarray}}
\newcommand{\ba}{\begin{array}}
\newcommand{\ea}{\end{array}}

\def\bbox{{\,\lower0.9pt\vbox{\hrule \hbox{\vrule height 0.2 cm
\hskip 0.2 cm \vrule height 0.2 cm}\hrule}\,}}
\newcommand{\dsl}{\pa \kern-0.5em /}

\newcommand{\dR}{I\!\!R}

\newcommand{\mysection}[1]{\section{#1}
            \setcounter{equation}{0}\setcounter{figure}{0}}

\def\5{\bar }  \def\6{\partial } \def\7{\tilde }
\def\8{\hat }

\newcommand{\beq}{\begin{equation}}
\newcommand{\eeq}{\end{equation}}

\newcommand{\bear}{\begin{eqnarray}}
\newcommand{\eear}{\end{eqnarray}}

\begin{document}

\begin{titlepage}
\vfill
\begin{flushright}
CERN-TH/2000-326\\
hep-th/0011003\\
\end{flushright}

\vfill

\begin{center}
\baselineskip=16pt
{\Large\bf Quasi-Particles in  Non-Commutative Field Theory\footnote{
Talk given at NATO ARW "Non-Commutative Structures in Mathematics and 
Physics", 
Kyiv 24-27 Sep. 2000}}
\vskip 0.3cm
{\large {\sl }}
\vskip 10.mm
{ Karl Landsteiner }\\
\vskip 1cm

{\small
 Theory Division CERN\\
CH-1211 Geneva 23, Switzerland}\\
email: Karl.Landsteiner@cern.ch
\end{center}
\vskip5cm
\par
\begin{center}
{\bf ABSTRACT}
\end{center}
\begin{quote}
After a short introduction to the UV/IR mixing in non-commutative
field theories we review the properties of scalar quasi-particles
in non-commutative supersymmetric gauge theories at finite temperature.
In particular we discuss the appearance of super-luminous wave
propagation.
\end{quote}
\end{titlepage}

\mysection{Introduction}

Given the experience of quantum mechanics 
it seems a rather natural idea that spacetime at very small distance-scales
might be described by non-commuting coordinates \cite{lan-sny,lan-ncbooks}. 
Keeping the example of
quantum mechanics in mind one is lead to write down a commutation relation
for the coordinates such as
\begin{equation}\label{lan-eq1}
[x^m,x^n] = i \theta^{mn}\,.
\end{equation}
In order to study quantum field theory on such non-commuting spaces 
it is useful to make some further simplifying assumptions,
in particular we will take $\theta^{mn}$ to be an element of the centre of the algebra
defined by (\ref{lan-eq1}).


A convenient way of thinking about non-commutativity is by deformation
of the product on the space of ordinary function. Using $\theta^{mn}$ as 
deformation parameter we define the so-called Moyal product (or star-product)
by
\begin{equation}\label{lan-eq7}
f(x) * g(x) := \lim_{y\rightarrow x} e^{\frac{i}{2} \theta^{mn}\partial^x_m \partial^y_n}
f(x) g(y) \,.
\end{equation}
In momentum space it takes the form
\begin{equation}\label{lan-eq6}
f(x) * g(x) = \int \frac{d^n k}{(2\pi)^n} \int \frac{d^n q}{(2\pi)^n}
\tilde{f}(k) \tilde{g}(q) e^{-i(k+q)x} e^{-\frac{i}{2} k_m \theta^{mn} q_n} \,.
\end{equation}
An immediate consequence is that we can always delete
one star under the integral because the additional terms by which the
Moyal product differs from the usual product are total derivatives thanks
to the antisymmetry of $\theta^{mn}$
\begin{equation}\label{lan-eq8}
\int f(x) * g(x) d^nx = \int \left(f(x) . g(x) + \frac{i}{2} \theta^{mn}\partial_m f(x) \partial_n g(x) + \cdots\right)\, d^nx\,.
\end{equation}
This furthermore implies cyclic symmetry under integral
\begin{equation}\label{lan-eq9}
\int f*g*h = \int f. g*h = \int g*h.f = \int g*h*f\, .
\end{equation}

We have now all the ingredients do start discussing field theory. 
Before doing so we will introduce one further simplification, 
namely we will assume
from that time is an ordinary commuting coordinate, i.e. 
$\theta^{m0}=0$. This has the advantage that we are still dealing
with a system with a finite number of time derivatives. Although a
canonical formalism for theories with an infinite number of time 
derivatives can be developed \cite{lan-quim}
it turns out that quantum field theory
on spaces with time-space non-commutativity are not unitary at the
one-loop level \cite{lan-jaume}\footnote{This applies to the 
time-like case, i.e. in all coordinate systems with $\theta^{mn}=const$
the commutator (\ref{lan-eq1}) involves the time coordinate.}.

\begin{figure}
\centerline{
\epsfig{file=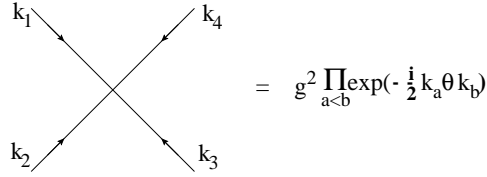,width=15pc} }
\caption{Feynman rule for non-commutative $\Phi^4$ vertex.}
\end{figure}
\begin{figure}
\centerline{
\epsfig{file=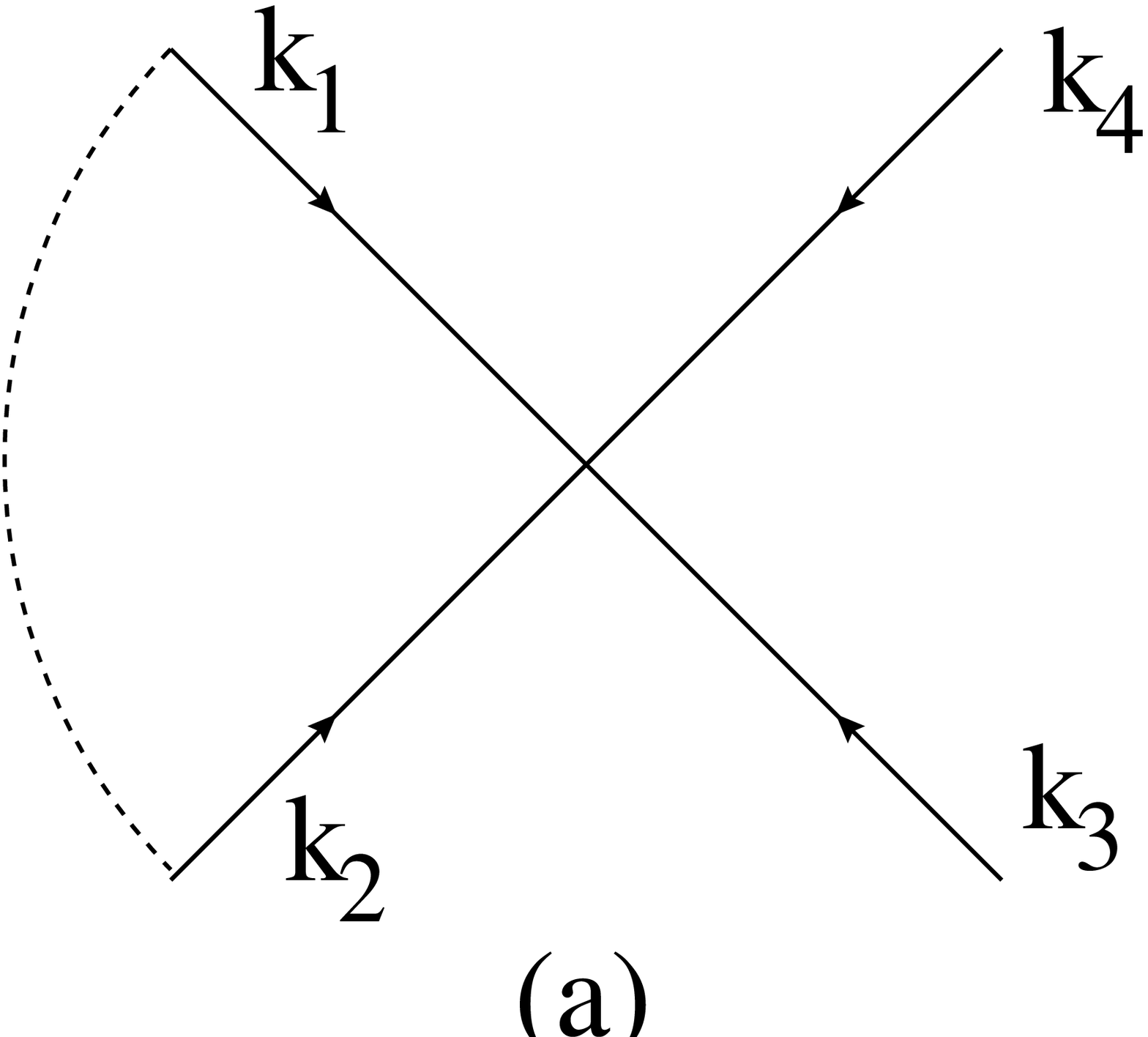,width=7pc}\qquad\qquad
\epsfig{file=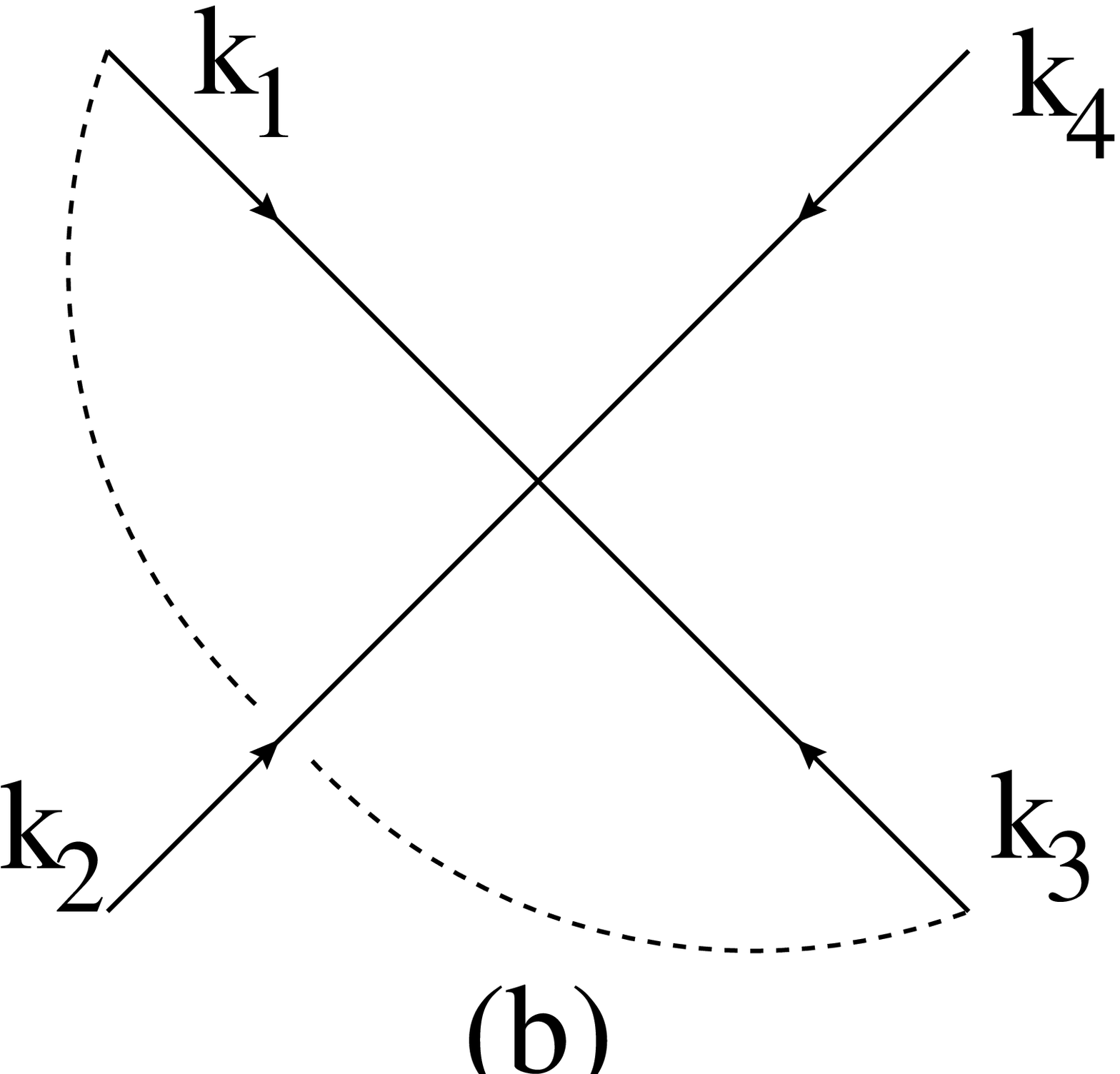,width=7pc}\qquad}
\caption{Corrections to the two-point function
can be either planar as in (a) or non-planar as in (b)}
\end{figure}

Non-commutative field theories can be viewed as non-local deformations
of local field theories. For fields of spin zero or
one-half we can take a Lagrangian of an ordinary field
theory and deform the product of fields according to the Moyal product 
(\ref{lan-eq8}), i.e. we replace the ordinary product by the star product.
For spin one-fields we also have to consider that the gauge symmetry is 
deformed, $\delta A_m = \partial_m \lambda +i \{A_m,\lambda\}_*$, 
where $\{.,.\}_*$
denotes the Moyal bracket $\{f,g\}_* = f*g - g*f$. The non-commutative
field strength of a gauge field is defined accordingly as $F_{mn}=\partial_m
 A_n - \partial_n A_m +i\{A_m,A_n\}_*$ and the covariant derivative as
$D_m . = \partial_m . + i\{A_m,.\}_*$ \cite{lan-cr}.

Let us consider now a scalar $\Phi^4$ theory on in four dimensions.
Without further loss of generality we assume $\theta^{23}=-\theta^{32}=\theta$.
Because one can drop one star-product in the Lagrangian the free theory
is unchanged with respect to the one on ordinary $\dR^n$. 
The tree level propagator is then the usual one
\begin{equation}\label{lan-eq10}
\langle \Phi(p) \Phi(-p) \rangle = \frac{i}{p^2-m^2}\,.
\end{equation}
The one-loop corrections to the two point function that arise from
the $\Phi^4$ vertex are shown in figure 2(a) and 2(b).
Because of the cyclic symmetry of the vertex we have two distinct classes
of graphs \cite{lan-gonz}. If we connect neighbouring lines of the vertex in figure (1)
the dependence of the exponential on the internal momentum $k=k_1=-k_2$
cancels. Thus the diagram 2(a) gives rise to a quadratic divergence 
in the same way as it happens in ordinary $\Phi^4$ theory.

If we contract however non-neighbouring lines the dependence on the
internal momentum of the exponent does not cancel. The distinct classes
of Feynman diagrams in non-commutative field theories are called 
planar if they are of type 2(a) and non-planar if they are of type 2(b).

The divergence is
regulated by the rapid oscillation of the exponential function at large
internal momentum and we find 
\begin{equation}\label{lan-eq11}
4 g^2 \int \frac{d^4k}{(2\pi)^4} \frac{e^{i\tilde{p}k}}{k^2-m^2} = \frac{i g^2}{4\pi^2 \tilde{p}^2 } +\cdots\,,
\end{equation}
Where we introduced the notation $\tilde{p}^n = p_m \theta^{mn}$ and
the dots indicate terms that are less singular for $\tilde{p}\rightarrow 0$.
Resummation gives rise to a corrected two-point function on the one loop level
of the form
\begin{equation}\label{lan-eq12}
\Gamma^2(p) = p^2 - m_R^2 + \frac{g^2}{ \pi^2 \tilde{p}^2} \,.
\end{equation}
The quadratic divergence in the planar graph gives rise to a 
renormalization of the mass. The non-planar graph results in a dramatic
change of the infrared behaviour of the theory. On a technical level 
the origin of this infrared divergence is easily understood. The non-planar
diagram is regulated by the phase factor stemming from the star product.
This phase is absent if the external momentum flowing into the diagram
vanishes. Thus the ultraviolet divergence has been converted into an
infrared divergence. This phenomenon UV/IR mixing has first been discussed in
\cite{lan-mrs} and has been further investigated in \cite{lan-haya}-
\cite{lan-adbs}. Notice also
that the IR-singularity is present even in the massive theory. Since it is
induced by modes in the far UV circling in the loop it is insensitive to the
presence of a massterm. 

It should be
emphasised that there are usually also sub-leading logarithmic infrared
divergences. In the infrared these become important at momenta of the order of
$p = {\cal O}(e^{-\frac{1}{g^2}})$. Down to these non-perturbatively small
momenta the infrared behaviour is dominated by the effects stemming from
the quadratic divergences. In the following we will always concentrate
on the leading order IR-behaviour and thus neglect the contributions from
the logarithms.

In supersymmetric theories quadratic divergences in four dimensions are
absent. 
However at finite temperature supersymmetry is broken and the one-loop 
dispersion relation will again show effects from UV/IR mixing in
non-planar graphs. Because temperature acts as a cutoff no
IR-singularities are to be expected. The next section reviews these
effects in the example of ${\cal N}=4$ supersymmetric Yang-Mills theory.

\mysection{Quasi-particles in non-commutative ${\cal N}\!=\!4$ SYM} 
\label{zeroT}
We limit ourselves to 
the study of a non-commutative $U(1)$ ${\cal N}\!=\!4$ gauge theory.
The spectrum of the theory consists of six scalars, four Majorana Fermions
and a vector field. The Lagrangian takes the form 
\begin{equation}\label{lan-eq21}
\begin{array}{rl}
{\cal L} =& \frac{1}{g^2}\int \left(  -\frac{1}{4} F_{mn} F^{mn} + 
\frac{1}{2}D_m\Phi^{ab} 
D^m \Phi_{ab} +\frac{1}{4} \{\Phi^{ab},\Phi^{cd}\}_* \{\Phi_{ab},\Phi_{cd}\}_*
\right.+\\
&\left. + i \lambda_a 
\sigma^m D_m\bar{\lambda}^a 
+i\{\lambda_a,\lambda_b\}_* \Phi^{ab}+
i\{\bar\lambda^a,\bar\lambda^b\}_* \Phi_{ab}\right)\,.
\end{array}\end{equation} 
The theory has a global $SU(4)$ symmetry under which the fermions transform
under the $4$, $\bar{4}$. The 6 scalars transform in the antisymmetric.
This symmetry is indicated by indices $a,b$.

We will study the dispersion relation
of the ${\cal N}\!=\!4$ scalars at finite temperature and one loop level.
Finite temperature is implemented in the Matsubara formalism by considering
the theory on $S^1\times\dR\times\dR^2_{nc}$. The last factor indicates
the two-dimensional non-commutative plane. The fermions are taken to have
anti-periodic boundary conditions on the $S^1$ factor. Non-commutative
field theories at finite temperature have been investigated in 
\cite{lan-avm}-\cite{lan-we}

The scalar self-energy is given by
\begin{equation}\label{lan-eq22}
\Sigma_T = 32 g^2 \int
 {d^3k \over (2\pi)^3} {\sin^2 {\tilde{p}\cdot k\over  2} \over k} 
\left( n_B(k) + n_F(k) \right) + 4 g^2 P^2 \bar{\Sigma} \,,
\end{equation}
 $ n_B(k)$ and $n_F(k)$ denote Bose-Einstein and 
Fermi-Dirac distributions. 
Four momentum is denoted by $P^2 = p_0^2 - p^2$, lowercase denotes
three-momentum. Momenta along the non-commutative directions as will 
be called transverse. 

The first term in (\ref{lan-eq22}) vanishes at $T=0$ because of 
supersymmetry.
The second term contributes 
to the finite temperature wave-function renormalization of the scalar field. 
It affects the position of the pole only to ${\cal O}(g^4)$ and we
will drop it in the sequel.

Using the relation $\sin^2{\tilde{p} k\over 2} = {1\over 2} (1 - \cos\tilde p k)$ we can separate the planar and non-planar contributions to the self-energy.
The dispersion relation becomes
\begin{equation}\label{lan-eq23} \omega^2 = p^2 + 2g^2 T^2 - 
{4 g^2 T\over \pi |\tilde{p}|} \tanh{\pi |\tilde{p}| T\over 2} \, .
\end{equation}

\begin{figure}
\centerline{
\epsfig{file=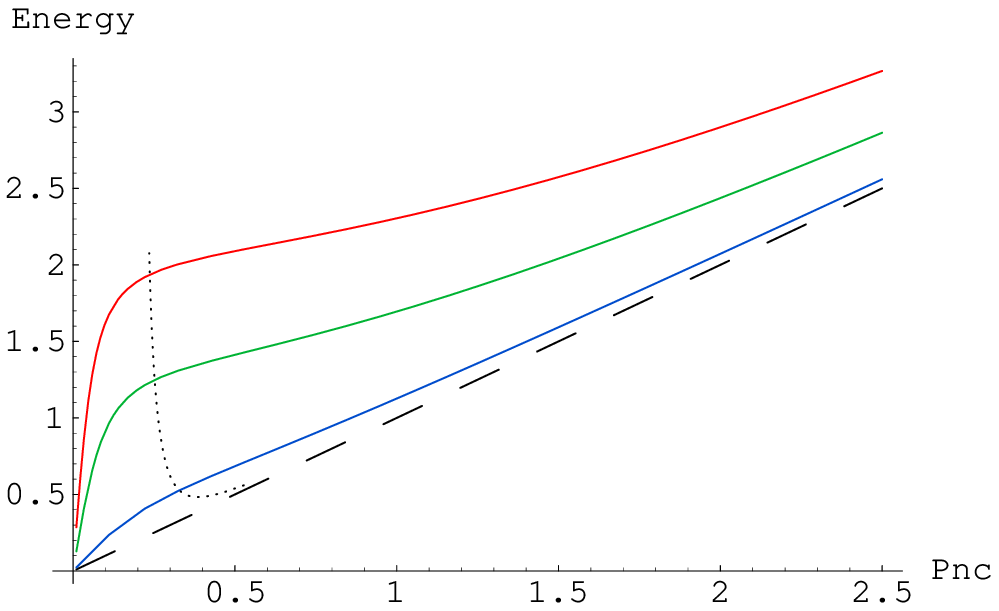,width=20pc} }
\caption{Dispersion relation for scalars in ${\cal N}\!=\!4$ Yang-Mills for different temperatures.
The momentum $p$ is taken to lie entirely in the non-commutative directions.
The dashed line shows the light cone $\omega=p$. The dotted line
shows the momentum $p_c$ below which the group velocity 
${\partial \omega\over \partial p}$ is bigger
than one.}
\end{figure}

A plot of the dispersion relation is shown in figure (3). 
The hyperbolic tangent arises solely from the non-planar contribution to the 
dispersion relation.

For large transverse external momenta the non-planar contribution
is sub-leading with respect to the planar one,
\begin{equation}\label{lan-eq24}
\omega^2 \approx p^2 + 2 g^2 T^2 - 4 g^2 {T\over \pi |\tilde p|}, \;\;\;\  T\tilde p \gg 1.
\end{equation}
The second term comes from the planar diagrams and gives a mass to the 
scalar excitations.
The sub-dominant term linear in $T$ arises solely from soft bosons in 
non-planar diagrams. 
These are modes with characteristic momentum $k \ll T$ and large occupation number
$n_B \approx T/k ~\gg~1$,
\begin{equation}\label{lan-eq241}
\Sigma_{np} \sim \int d^3k {1\over k} \cos{\tilde p  k} {T\over k} \sim {T\over \tilde p} \, .
\end{equation}

In usual space-time the approximation $n_B \approx T/k ~\gg~1$ results in
the well known ultraviolet catastrophe of classical field theory.
In the non-planar sector of non-commutative space this does not happen as 
long as $\tilde p$ is different from zero.
This is yet another 
manifestation of the UV/IR mixing of non-commutative field theories: to
leading order at high temperature, the non-planar contribution is effectively purely classical \cite{lan-fgkmpp}. 

At low transverse external momenta, the non-planar contribution tends to 
cancel 
the planar one.
For zero external transverse momentum the interaction switches off.
The theory
becomes a free, gap-less $U(1)$ gauge theory with $\omega^2 \approx p_3^2$.

Let us consider now the case where the momentum lies
along the non-commutative directions.
Since $\omega(0)=0$ and for large $p$, $\omega(p) \approx  \sqrt{p^2+2 g^2 T^2}$, which lies above the lightcone, 
there is a region in between with ${\partial\omega(p)\over \partial p} > 1$.
 Thus the group velocity must exceed the speed of light
for small transverse momenta! 
\begin{equation}\label{lan-eq25}
\omega^2 \approx \left(1 + {g^2 \pi^2 T^4 \theta^2\over 6} \right)\; p^2 \, .
\end{equation}
The low momentum excitations are massless, but propagate with an index of refraction $n = p/\omega$ 
 smaller than one. Because the interactions switch off at low momenta, 
we expect these modes to be long-lived. 
In figure (3) the momentum $p_c$ below which the group velocity exceeds
one is depicted by a dotted line. The dashed line shows the light cone 
$\omega=p$.

Let us  emphasise that these qualitative features should be quite general and not an artifact of our one loop approximation,
as they simply arise from the fact that the theory is non-interacting at zero transverse momentum and develops a 
mass gap otherwise\footnote{One might also be worried if these effects are
gauge dependent. A model without gauge symmetry can be obtained if one
sets the gauge field and one fermion
(the field content of an ${\cal N}\!=\!1$ vector multiplet)to zero.
This would result in a ${\cal N}\!=\!1$ Wess-Zumino model
with Moyal bracket interactions. It would only change the overall factor
in (\ref{lan-eq22}).}.

We now investigate  the consequences of the  dispersion relation 
(\ref{lan-eq25}) for wave propagation.
Imagine that some disturbance of the scalar field is created in the thermal bath at time $t=0$.   
To simplify matters we will consider only a one dimensional 
problem with momentum pointing in a non-commutative direction.
The fastest moving modes are the ones with longest wavelength. These are also 
the modes 
which are long lived in the thermal bath. For these
 it is possible to obtain the exact asymptotic behaviour by noting that
the dispersion relation around $k=0$ is 
\begin{equation}\label{lan-eq26}
\omega(k) = c_0 k - \gamma k^3 + O(k^5)\,,
\end{equation}
with $c_0 = \sqrt{1 + {g^2 \pi^2 T^4 \theta^2 \over 6}}$ and $\gamma = {g^2 \pi^4 \theta^4 T^6\over 120 c_0}$.
This is the dispersion relation of the linearised Korteweg-deVries equation 
whose solution
is expressed in terms of the Airy function $Ai(z)$. We can express the 
solution
for the head of a wavetrain by \cite{lan-whit}
\begin{equation}\label{lan-eq27} 
\Phi = {A \over 2(3\gamma t)^{1\over3}} Ai\left(
{x-c_0 t\over (3\gamma t)^{1\over3}}\right) \,.
\end{equation}
The Airy function has oscillatory behaviour for negative argument
and decays exponentially for positive argument. Thus the wavetrain decays
exponentially ahead of $x=c_0 t$. Behind the wave becomes oscillatory.
In this region one can match the Airy function with the asymptotics obtained
from a stationary wave approximation. 
In between the oscillatory region and the exponential decay
there is a transition region of width proportional to $(\gamma t)^{1\over 3}$
around $x=c_0 t$. In this region the wavetrain has its first crest which
therefore is moving with a velocity approximately given by $c_0$.

Group velocities faster than the speed of light do also appear in
conventional physics, e.g. it is well-known that this happens
for light propagation in media close to an absorption line. 
Since the dispersive effects are however large, 
the group velocity loses its meaning as the velocity of signal
transportation. In our case, it is interesting to notice that
as the temperature increases, not only $c_0$ but also $\gamma$ grows.
This implies that at high temperatures the soft transverse momenta
become very dispersive. In such situations it is useful to introduce the 
concept of a front
velocity which is the velocity of the head of the wavetrain.
For the propagation of light in a medium it can be shown
that this front velocity never exceeds the speed of light even if
the group velocity can be faster than the speed of light \cite{lan-brill}. 
In our case the front velocity can be defined as the velocity of
the first crest of the wavetrain. According to (\ref{lan-eq25}) and
(\ref{lan-eq26}) this is always bigger that the speed of light.
The advance of the first crest with respect to an imagined
light front is $(c_0-1)t$. Since its spread grows as $(\gamma t)^{1
\over 3}$, the first crest is well defined outside the lightcone
for large enough time, $t> t_0$ where $t_0=\sqrt{\gamma \over (c_0-1)^3}$.

The question arises if this super-luminosity implies a violation of
causality. This is not necessarily the case. Violation of causality 
needs both ingredients: super-luminosity and the relativity principle.
Imagine an observer A emitting some signal with super-luminous
velocity $c_0$. If the relativity principle is valid another observer
B in a boosted frame relative to A could then catch the signal. B could
send an answer also with super-luminous speed $c_0$. 
The answer would reach observer A before he sent out
the original signal. The crucial point is of course that in the
non-commutative space-time we are considering boosts are not anymore
symmetries. In particular only in the frame of observer A time
is ordinary, commuting time. Any other frame involving a boost in
a non-commutative direction implies that also time is non-commutative.
To obtain an answer if causality is violated one would have to
calculate the dispersion relation also in such a frame and study
wave propagation then. Finite temperature field theory with
non-commutative time is however difficult to formulate due to
the infinite number of time derivatives appearing in the star product.
This is an open question though progress could possibly be achieved along
the lines in \cite{lan-quim}.

\mysection{Discussion and Outlook}
We have concentrated on reviewing the properties of scalar quasi-particles
at finite temperature in non-commutative ${\cal N}\!=\!4$ gauge theory.
Another system that has been studied in \cite{lan-we} is the non-commutative
Wess-Zumino model with star-product interactions instead of Moyal-brackets.
The one-loop self-energy is given by a similar expression as 
(\ref{lan-eq22}) except that $\sin^\frac{\tilde{p}k}{2}$ is substituted by
$\cos^\frac{\tilde{p}k}{2}$. It turns out that
this has the effect that for temperatures $T>T_0\approx\frac{1}{\sqrt{g\theta}}$
the minimum
of the dispersion relation is displaced from $p=0$! It has been argued that
this makes Bose-condensation of scalar modes impossible for temperatures
higher than $T_0$ \cite{lan-we} \footnote{The phase structure of a non-commutative
scalar field model in four dimensions has been investigated in \cite{lan-gubs}
where it has been argued that condensation to stripe phases occurs.}. 

Another system that has been studied in \cite{lan-we} was ${\cal N}\!=\!2$ gauge
theory at finite density. The results are qualitatively analogous to the
case with temperature. The role of the temperature is then played by
the chemical potential. 

Non-commutative field theories in the setup discussed here appear
also in string theory. In \cite{lan-sw} it was shown that the physics of
D-branes in a $B$-field background in a particular scaling limit with
$\alpha^\prime\rightarrow 0$ is described by non-commutative supersymmetric
gauge theories. It has been suggested that the effects of UV/IR mixing
could be understood from a string perspective \cite{lan-mrs}. 
The UV/IR mixing in this stringy context has been considered in
\cite{lan-kl}-\cite{lan-crs}.
Since the model considered here arises as the scaling limit of 
a D-3-brane in a $B$-field background it would be very interesting 
to reconsider the one-loop dispersion relations from a string 
theory perspective.

\end{document}